\documentclass{aa}  

\usepackage{graphicx}
\usepackage{txfonts}
\usepackage{newtxmath}

\newcommand{\vect}[1]{\boldsymbol{#1}}
\newcommand{\HI}{\text{H}\textsc{i}}

\begin{document}

   \title{Degree-Scale Galactic Radio Emission at 122\,MHz around the North Celestial Pole with LOFAR-AARTFAAC}

   \author{B. K. Gehlot\inst{1}\fnmsep\thanks{kbharatgehlot@gmail.com},
           L. V. E. Koopmans\inst{1},
           A. R. Offringa\inst{2,1},
           H. Gan\inst{1},
           R. Ghara\inst{3}
           S. K. Giri\inst{4}
           M. Kuiack\inst{5},  
           F. G. Mertens\inst{6,1}, 
           M. Mevius\inst{2},
           R. Mondal\inst{7}
           V. N. Pandey\inst{2,1},
           A. Shulevski\inst{8,5,2}
           R. A. M. J. Wijers\inst{5}, 
           and S. Yatawatta\inst{2}
          }

   \institute{Kapteyn Astronomical Institute, University of Groningen, PO Box 800, 9700AV Groningen, The Netherlands.
         \and
              ASTRON, Netherlands Institute for Radio Astronomy, Oude Hoogeveensedijk 4, 7991 PD, Dwingeloo, The Netherlands.
         \and
              Department of Physics, Technion, Haifa 32000, Israel
         \and
              Institute for Computational Science, University of Zurich, Winterthurerstrasse 190, 8057 Zurich, Switzerland.
         \and
              Anton Pannekoek Institute, University of Amsterdam, Postbus 94249 1090 GE, Amsterdam, The Netherlands.
         \and
              LERMA, Observatoire de Paris, PSL Research University, CNRS, Sorbonne Universit\'e, F-75014 Paris, France.
         \and
              The Oskar Klein Centre, Department of Astronomy, Stockholm University, AlbaNova, SE-10691 Stockholm, Sweden.  
         \and
              Leiden Observatory, Leiden University, PO Box 9513, NL-2300 RA Leiden, The Netherlands.
             }

   \titlerunning{Degree-Scale Galactic Foregrounds}
    \authorrunning{B. K. Gehlot et al.}

   \date{Received ---; accepted ---}

  \abstract
   {}
   {Contamination from bright diffuse Galactic thermal and non-thermal radio emission poses crucial challenges in experiments aiming to measure the 21-cm signal of neutral hydrogen from the Cosmic Dawn (CD) and Epoch of Reionization (EoR). If not included in calibration, this diffuse emission can severely impact the analysis and signal extraction in 21-cm experiments. We examine large-scale diffuse Galactic emission at 122~MHz, around the North Celestial Pole, using the Amsterdam-ASTRON Radio Transient Facility and Analysis Centre (AARTFAAC)- High Band Antenna (HBA) system.}
   {In this pilot project, we present the first-ever wide-field image produced with a single sub-band of the data recorded with the AARTFAAC-HBA system. We demonstrate two methods: multiscale CLEAN and shapelet decomposition, to model the diffuse emission revealed in the image. We use angular power spectrum metrics to quantify different components of the emission and compare the performance of the two diffuse structure modelling approaches.}
   {We observe that the point sources dominate the angular power spectrum ($\ell(\ell+1)C_{\ell}/2\piup \equiv \Delta^2(\ell)$) of the emission in the field on scales $\ell\gtrsim 60$ ($\lesssim 3$~degree). The angular power spectrum  after subtraction of compact sources is flat within $20\lesssim \ell \lesssim200$ range, suggesting that the residual power is dominated by the diffuse emission on scales $\ell\lesssim200$. The residual diffuse emission has a brightness temperature variance of $\Delta^2_{\ell=180} = (145.64 \pm 13.61)~{\rm K}^2$ at 122~MHz on angular scales of 1~degree, and is consistent with a power-law following $C_{\ell}\propto \ell^{-2.0}$ in $20\lesssim \ell \lesssim200$ range. We also find that, in the current setup, the multiscale CLEAN is suitable to model the compact and diffuse structures on a wide range of angular scales, whereas the shapelet decomposition method better models the large scales, which are of the order of a few degrees and wider.}
   {}


   \maketitle
%

\section{Introduction}

Observations of the redshifted 21-cm hyperfine transition line of neutral hydrogen ($\HI$ hereafter) are expected to unveil the properties of the first luminous objects and the evolution of the large-scale structure during the Cosmic Dawn (CD; $12\lesssim z\lesssim 30$) and Epoch of Reionization (EoR; $6\lesssim z\lesssim 12$) \citep{madau1997,shaver1999,furlanetto2006}. Several current and next-generation experiments such as GMRT \citep{paciga2013}, LOFAR \citep{vanhaarlem2013}, MWA \citep{tingay2013,bowman2013}, NenuFAR \citep{zarka2012,mertens2021}, OVRO-LWA \citep{eastwood2019}, AARTFAAC \citep{gehlot2020}, HERA \citep{deboer2017}, SKA \citep{koopmans2015}, EDGES \citep{bowman2018}, SARAS3 \citep{nambissan2021}, LEDA \citep{bernardi2016}, and REACH \citep{deleraacedo2019} are attempting to measure the fluctuations and global evolution of the 21-cm brightness temperature as a function of redshift.

However, the faint 21-cm signal from high redshifts is contaminated by astrophysical foreground emission that is 3-4 orders of magnitude brighter than the signal of interest. The astrophysical foregrounds consist of Galactic diffuse emission (synchrotron and free-free emission), and extra-galactic compact sources such as radio-galaxies, supernova remnants and other sources \citep{dimatteo2002,zaldarriaga2004,bernardi2009,bernardi2010,ghosh2012}. Subtraction or avoidance of these bright foregrounds poses a significant challenge in all 21-cm cosmology experiments. The interferometric experiments such as PAPER and HERA follow the so-called ``foreground-avoidance'' technique for separating the 21-cm signal from the foregrounds. This technique takes advantage of the fact that the spectrally smooth foregrounds occupy a small number of spectral modes and tend to reside within a ``wedge-shaped'' region of the Fourier space, whereas the ``EoR-window'', in contrast to the ``wedge'' region, should be dominated by 21-cm signal and the noise \citep{datta2010,morales2012,vedantham2012,trott2012,parsons2012,hazelton2013,dillon2014,morales2019}. On the other hand, 21-cm experiments with LOFAR, AARTFAAC, NenuFAR, and also the planned SKA follow the foreground-removal approach in their analysis strategy \citep{gehlot2019,gehlot2020,mertens2020}. The 21-cm experiment with MWA utilises foreground subtraction in conjunction with foreground avoidance \citep{barry2019,trott2020}. Current foreground removal techniques used by experiments in the latter category include image deconvolution and source extraction algorithms such as CLEAN \citep{hogbom1974,clark1980,cornwell2008,offringa2017}, Duchamp \citep{whiting2012}, and PyBDSF \citep{mohan2015} to model the extra-galactic compact sources as delta functions and Gaussians in image space. This is followed by subtraction of the model from the observed signal using Direction Dependent (DD) calibration, e.g., with \textsc{sagecal} \citep{yatawatta2016,yatawatta2017}. However, these traditional methods, largely developed for longer-baseline high brightness-temperature data, are sub-optimal for modelling large-scale diffuse emission, extended sources and foreground emission below the confusion noise that dominates the observed signal. The latter are typically mitigated in these experiments using blind foreground removal techniques such as Wp Smoothing \citep{harker2009}, FastICA \citep{chapman2012}, GMCA \citep{chapman2013}, and GPR \citep{mertens2020} that exploit the spectral smoothness of the foregrounds to model and remove foregrounds from the observed data. 

Besides the intrinsic interest of studying properties of the diffuse Galactic emission, it is essential to include the diffuse (and compact) emission component in the instrumental gain calibration step regardless of the signal separation choices adopted by various experiments. Various investigations by \cite{patil2016}, \cite{barry2016}, \cite{ewall-wice2017}, \cite{sardarabadi2019}, and \cite{byrne2019} showed that using an incomplete sky-model in the calibration step leads to significant suppression of the diffuse emission as well as the 21-cm signal of interest on shorter baselines that are dominated by diffuse emission. Since shorter baselines provide most of the sensitivity toward the 21-cm signal, it is crucial to calibrate these baselines accurately to prevent any signal suppression. One option to mitigate the effect of sky-incompleteness is to remove the short baselines from calibration and include only longer baselines where the diffuse emission is resolved out \citep{patil2017}. However, the exclusion of short baselines from calibration leads to a so-called excess variance on the excluded baselines in calibration \citep{patil2016,barry2016,ewall-wice2017,sardarabadi2019}, impacting the power spectrum estimation. Data from several 21-cm experiments such as MWA \citep{byrne2021}, OVRO-LWA \citep{eastwood2018} and LWA New Mexico station \citep{dowell2017} are already being used to develop diffuse sky maps to facilitate precise calibration of 21-cm observations. As a part of the ACE project \citep{gehlot2020}, the data from AARTFAAC-LBA is also being utilised to produce a wide-band map of the northern sky ($\sim$10~arcmin resolution) at low frequencies with high fidelity due to its superb $uv$-coverage.    

In this work, we present the first-ever wide-field images produced using the High-Band Antenna (HBA) observations of the LOFAR Amsterdam-ASTRON Radio Transients Facility And Analysis Centre (AARTFAAC) wide-field imager \citep{prasad2016} at 122~MHz. We explore multiscale CLEAN and shapelet decomposition methods to model the diffuse emission in an extended region of around $20^{\circ}$ radius centred at the North Celestial Pole (NCP). \cite{line2020} also presented a similar study comparing multiscale CLEAN and shapelet decomposition to model the bright extended source Fornax~A using simulations and data from MWA. However, that study was focused on modelling sources with extended emission from a few arcmins up to a degree. In this pilot project, we focus on quantifying the spatial properties of the spectrally smooth diffuse emission in the field using the angular power spectrum technique. In the future, we plan to expand the current analysis by including spectral and polarization information of the diffuse emission. 

The paper is organised as follows: section~\ref{sec:obs_preprocess} and~\ref{sec:cal_img} briefly describe the observational setup, preprocessing steps, calibration, and imaging scheme. The two diffuse foreground modelling methods and their comparison are presented in section~\ref{sec:modeling}. Section~\ref{sec:angular_PS} describes the angular power spectrum to quantify the properties of the diffuse emission. Finally, section~\ref{sec:summary_futurework} provides a summary of the work and lays out the future steps and improvements.

\begin{table}
	\centering
	\caption{Observational details of the data.}
	\label{tab:obs_details_A12HBA}
	\begin{tabular}{ll}	
		\hline		
		\textbf{Parameter} & \textbf{value} \\	
		\hline
		Telescope & LOFAR-AARTFAAC \\
		Antenna configuration & \texttt{A12} \\
		Number of receivers & 576 (HBA tiles) \\
        Number of pointings & 1 \\
		Pointing centre (RA, Dec; J2000): & 0h0m0s, $+90^{\circ}00^{\prime}00^{\prime\prime}$ \\
		Duration of observation & 11 hours   \\
        Observing frequency (and width) & 122~MHz (195.3~kHz)\\
		Primary Beam FWHM & $28^{\circ}$ at 122~MHz \\
		Field of View &  $625\,\text{deg}^2$ at 122~MHz\\	
		Polarisation & Linear X-Y   \\
		Time, frequency resolution: \\
		\quad Raw Data & 2 s, 195.3 kHz      \\
		\quad After flagging & 12 s, 195.3 kHz \\
		\hline
	\end{tabular}
\end{table}

\section{Observations and preprocessing}\label{sec:obs_preprocess}

We use the AARTFAAC-HBA wide-field imager to observe an extended region of $\sim$20-degree radius around the NCP in the frequency range of $114-126$\,MHz, which is the primary observation window of the LOFAR-EoR Key Science Project (KSP) \citep{patil2017,mertens2020}. The observational setup and the preprocessing steps are discussed briefly in the following subsections.

\subsection{The AARTFAAC Wide-Field Imager}\label{subsec:A12_HBA}

AARTFAAC is a LOFAR-based all-sky radio transient monitor \citep{kuiack2019}. It piggybacks on ongoing LOFAR observations and taps the digital signal streams of individual antenna elements from six (\texttt{A6} mode that uses ``superterp'' stations) or twelve core stations (\texttt{A12} mode that uses 12 central stations). The \texttt{A6} mode consists of 288 dual-polarization receivers (e.g. Low Band Antenna (LBA) dipoles or HBA tiles) within a 300 m diameter circle, and the \texttt{A12} mode consists of 576 such receivers spread across 1.2 km diameter. The digitised signals from these receivers are tapped and transported to the AARTFAAC correlator prior to beam-forming. Due to current network capacity limitations, only 16 sub-bands can be correlated in the 16-bit mode. Each sub-band is 195.3 kHz wide and consists of up to 64 channels providing a maximum frequency resolution of 3 kHz, with an instantaneous system bandwidth of 3.1 MHz. The correlator subsystem, located at the Centre for Information Technology (CIT) at the University of Groningen (the Netherlands), is a GPU based correlator that maximally handles 576 input signal streams per polarization (1152 total streams) and produces XX, XY, YX, YY correlations for all receiver pairs for every frequency channel with 1-second integration. The output correlations can either be dumped as raw correlations on disks or routed to the AARTFAAC real-time calibration and imaging pipeline for transient detection \citep{prasad2016,kuiack2019}. Readers are referred to \cite{vanhaarlem2013} and \cite{prasad2016} for detailed information about LOFAR and AARTFAAC system design and observing capabilities.

\begin{figure}
    \centering
    \includegraphics[width=1\columnwidth]{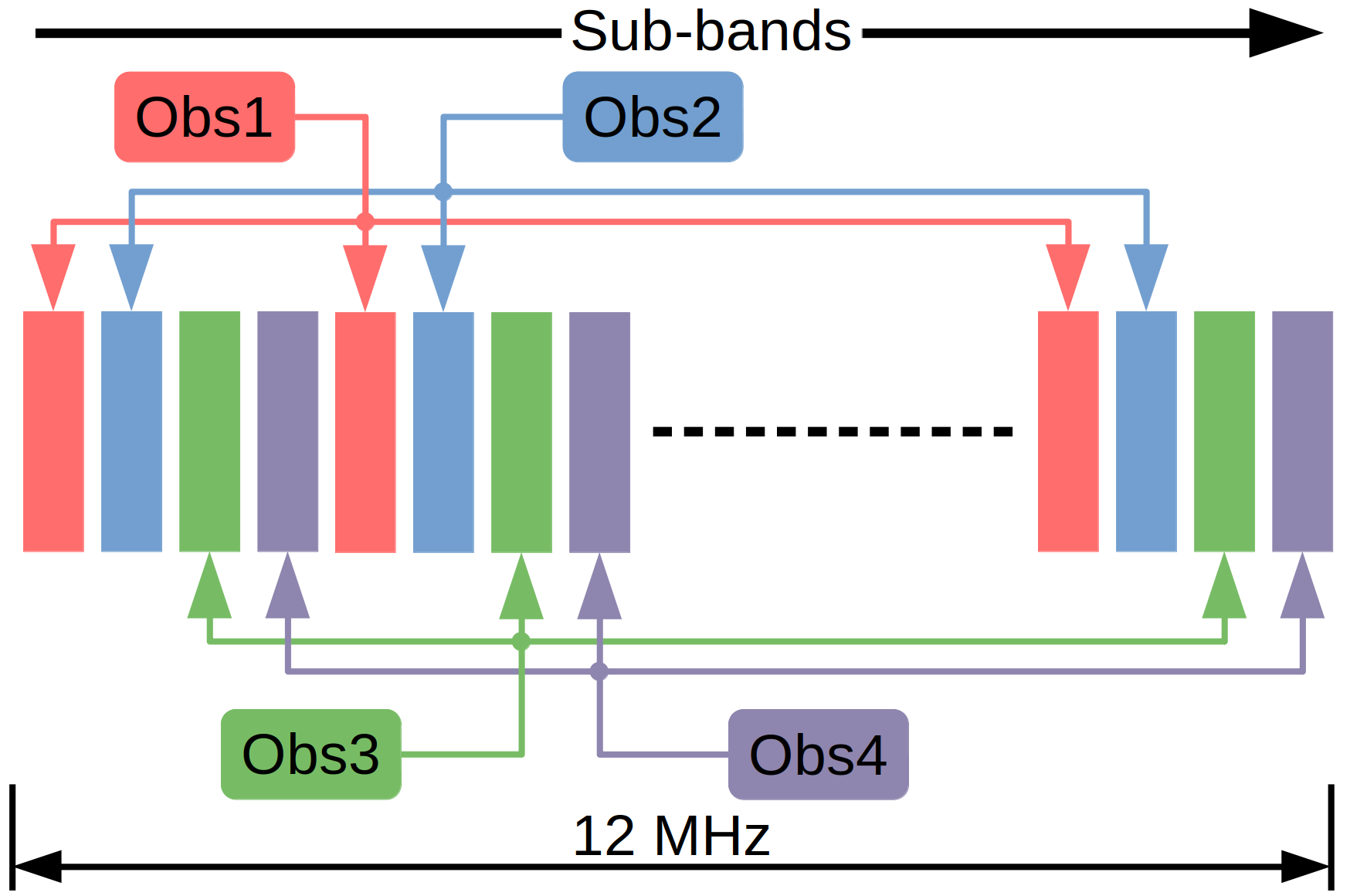}
    \caption{Schematic of the frequency comb configuration utilised for the observations. The coloured blocks correspond to sub-bands spanning the 12~MHz bandwidth. Each colour represents sub-bands corresponding to the same observation.}
    \label{fig:SB-config}
\end{figure} 

\subsection{Observational setup}\label{subsec:obs_setup_A12HBA}

We use the AARTFAAC-HBA system in \texttt{A12} mode (A12-HBA hereafter) to observe the field around the NCP. For our observations, we target the 114-126\,MHz frequency range corresponding to the redshift range of $z = 10.2 - 11.4$. Due to the currently limited bandwidth of 3.1 MHz, we adopt an observation strategy where we combine four different observations recorded within a span of a few weeks to cover the 12\,MHz band. For each observation, sub-bands are sparsely spread in a frequency-comb configuration spanning the 12~MHz band. Sub-bands from each observation are interlaced and do not overlap. This observing strategy allows each observation to independently (and sparsely) span the 12~MHz bandwidth and continuous coverage of the 12~MHz band when combined together. Figure~\ref{fig:SB-config} shows a schematic of the sub-band setup that we have adopted for the observations. Although higher time and frequency resolution is preferred for RFI-flagging, we restricted the data resolution to lower values of  2~s and 195~kHz (to observe the NCP for a longer duration) due to the limited capacity of the AARTFAAC storage system. The raw correlation data is later transported to the LOFAR-EoR KSP processing system (Dawn HPC cluster, \citealt{pandey2020}), where all of the data processing and analysis are performed. The raw data is converted to a standard Measurement Set (MS) format using \textsc{aartfaac2ms}\footnote{\url{https://github.com/aroffringa/aartfaac2ms}}\citep{offringa2015} and the raw visibilities are phased to the NCP before preprocessing. The observational details of this dataset are summarised in table \ref{tab:obs_details_A12HBA}. 

\begin{figure*}
    \centering
    \includegraphics[width=0.85\textwidth]{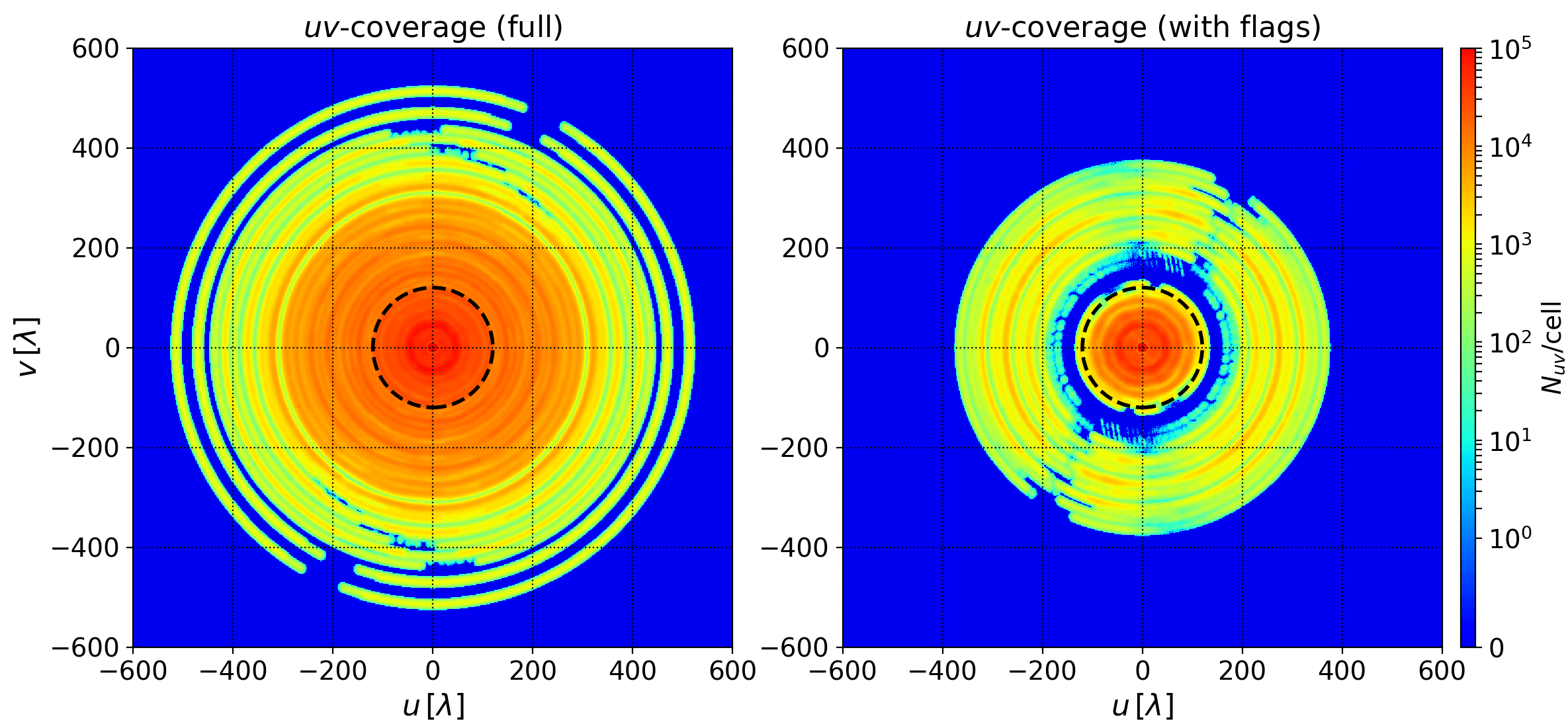}
    \caption{Single sub-band (122~MHz)  $uv$-coverage of the A12-HBA array towards the NCP for 11 hours of synthesis. Left Panel: $uv$-coverage of the full A12-HBA array. Right Panel: same as left but with flags included. The colour-scale corresponds to the number of baselines in $uv$-cells of size $(\delta u, \delta v) = (1,1)$ (in units of wavelength). The dotted black circle shows $|\vect{u}| = 120$, which is used as the baseline range to model the diffuse structure in later sections. The effect of flagging five~stations is clearly visible in the $uv$-coverage shown in the right panel.}
    \label{fig:uv-coverage}
\end{figure*}

\subsection{Data Preprocessing}\label{subsec:preprocess}

The preprocessing steps include RFI-flagging and averaging of the raw visibilities. We use \textsc{aoflagger} \citep{offringa2010,offringa2012} to flag RFI-corrupted data. We also flagged all the visibilities corresponding to non-operational tiles in the datasets. The remaining data were averaged to a lower time resolution of 12 seconds. After averaging, the data volume of an 11-hour observation of 3.1~MHz bandwidth is $\sim1.7$~TB. We noted that 5 out of 6 stations outside the ``superterp" had significantly lower visibility amplitudes in all observations recorded during this first pilot program. Furthermore, the visibilities corresponding to these stations showed extremely erratic behaviour even after calibration. Therefore, we also flagged the tiles corresponding to these five stations before the calibration step. Figure~\ref{fig:uv-coverage} shows the $uv$-coverage of the full A12-HBA array and after flagging the above stations. There is a significant reduction of long baselines due to the flagging of the five~stations. Note that a single sub-band is sufficient for the purpose of current analysis, i.e. to model and study the spatial properties of the spectrally-smooth diffuse foreground emission, as the confusion noise level is reached in a single sub-band. The analysis of the full frequency range, including a power-spectrum analysis, will be presented in a future publication. 

\begin{table*}
	\centering
	\caption{List of sources in the initial calibration model.}
	\label{tab:cal_model}
	\begin{tabular}{lll}		
		\hline		
		\textbf{Source} & \textbf{Intrinsic Flux} ($F_{\text{int}}$) & \textbf{Reference} \\	
		\hline
		3C61.1 & $38$~Jy (151.5~MHz) & \cite{baldwin1985} \\
		3C220.3 & $38$~Jy (150~MHz) & \cite{cohen2007}\\
		LQAC 244+085 001 & $6.2$~Jy (151.5~MHz) & \cite{baldwin1985}\\
		NVSS J011045+873822 & $5.1$~Jy (151.5~MHz) & \cite{baldwin1985}\\
		NVSS J190350+853648 & $4.9$~Jy (151.5~MHz) & \cite{baldwin1985}\\
		NVSS J062205+871948 & $4.9$~Jy (151.5~MHz) & \cite{baldwin1985}\\
		\hline
	\end{tabular}
\end{table*}

\section{Calibration and imaging}\label{sec:cal_img}

Visibilities measured by a radio interferometer are corrupted by errors due to instrumental imperfections and environmental effects. These corruptions are broadly classified into two categories, viz. Direction Independent (DI) effects such as complex receiver gains, global band-pass and a global ionospheric phase, and Direction Dependent (DD) effects that change with the apparent direction of the incoming electromagnetic signals due to the antenna voltage patterns (and consequently the tile beam), ionospheric phase fluctuations and Faraday rotation. Calibration refers to estimating the gains corresponding to these effects to obtain a reliable estimate of the true visibilities after the gain correction. We perform gain calibration in a self-calibration manner using the following steps:

\begin{enumerate}

 \item \label{step:sub} The first step is to remove the bright sources Cassiopeia~A (Cas\,A) and Cygnus~A (Cyg\,A) (with the apparent flux of several hundred Jansky) that superpose significant PSF (Point Spread Function) side-lobes over the field. We use the DD-calibration option within \textsc{sagecal-co} \citep{yatawatta2015,yatawatta2016,yatawatta2017} to subtract these sources from the raw visibilities. \textsc{sagecal-co} regularises the gain solutions with a spectrally smooth function in the gain calibration to improve the calibration quality. We use the Cas\,A and Cyg\,A shapelet model adopted from the LOFAR-EoR calibration sky-model to subtract these sources using their inferred directional gains. We use a solution interval of 2 minutes and 5 ADMM (Alternating Direction Method of Multipliers) iterations with a regularisation factor $\rho=5$ \citep{yatawatta2016}. We exclude $|\vect{u}|<10$ (in units of wavelength) baselines from the calibration\footnote{The baseline cut is used only during the fitting step of DD-calibration. The subtraction step includes all baselines.} to avoid most of the large-scale diffuse emission biasing the calibration but note that the diffuse emission extends over a much wider range of baselines as shown in later sections. At the current sensitivity level, we do not expect biases due to unmodeled diffuse emission on longer baselines overwhelming the calibration.  
  
 \begin{figure*}
    \centering
    \includegraphics[width=0.85\textwidth]{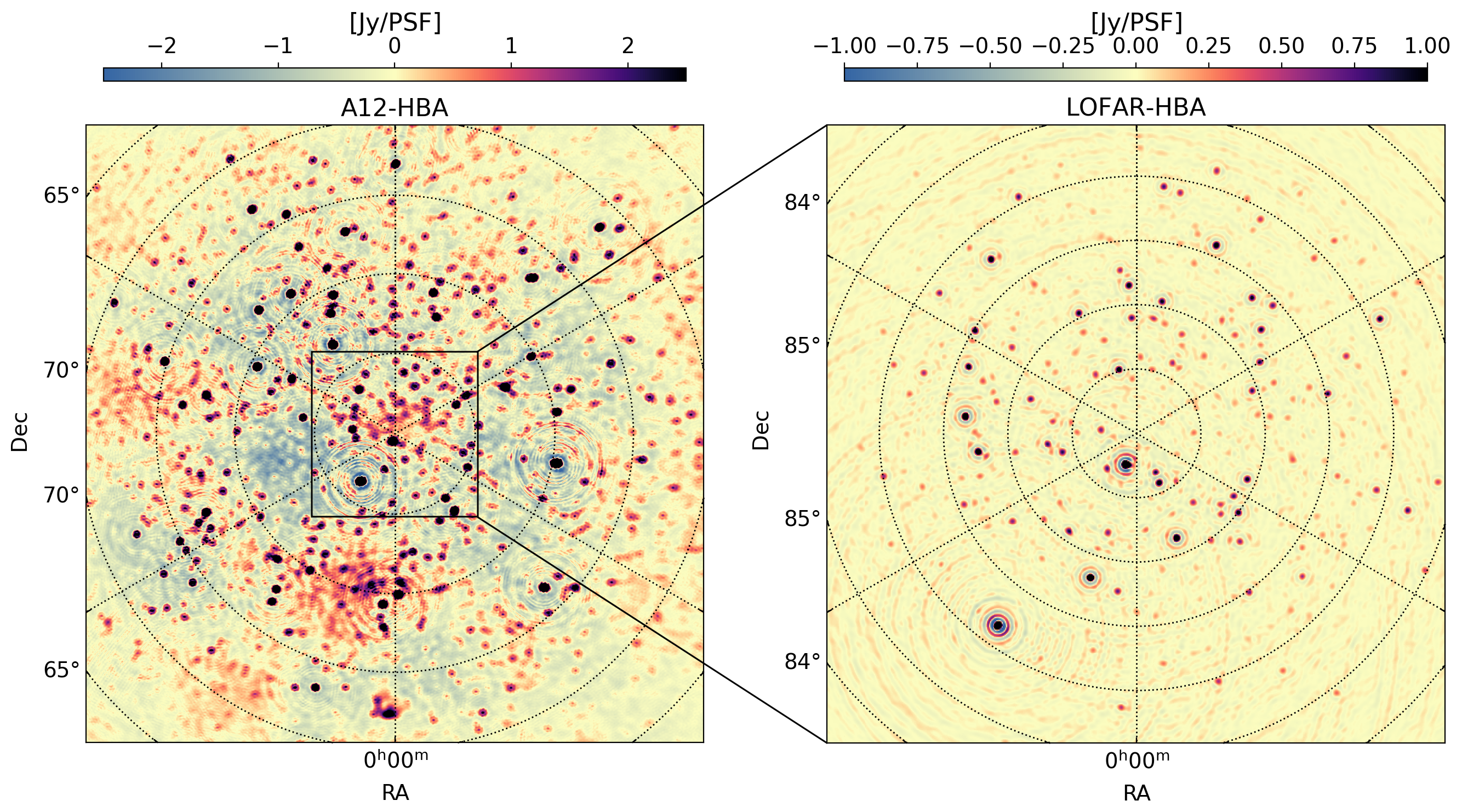}
    \caption{The left panel shows the Stokes $I$ dirty image of a $\sim$20\,degree radius field around the NCP using a single sub-band (122 MHz) of A12-HBA data and all A12-HBA baselines ($|\vect{u}|<400$) available after flagging. The right panel shows the Stokes $I$ image for the same field with the same imaging parameters and baseline range but using LOFAR-HBA station beam-formed data that lacks the shortest baselines. The dotted circles and spokes represent different declinations and right ascension, respectively.}
    \label{fig:dirty_compare}
\end{figure*} 
  
 \item \label{step:DIcal} We perform DI-calibration on the resulting visibilities after step \ref{step:sub}. The initial calibration model consists of six sources listed in table~\ref{tab:cal_model}, with intrinsic flux values ($F_{int}$) taken from \cite{cohen2007} (for 3C220.3) and \cite{baldwin1985} (remaining sources). All sources, except for 3C61.1, are represented by delta functions and a power-law representing their corresponding source spectra with an assumed but a typical spectral index of $-0.8$. The radio galaxy 3C61.1 has a more complicated morphology, and its model is adapted from the intrinsic sky-model used in the LOFAR-EoR calibration pipeline \citep{mertens2020}. The 3C61.1 model uses delta functions and shapelets to represent the source, and a third-order log-polynomial represents the spectrum. We use a calibration solution interval of 2 minutes to maintain a significant Signal-to-Noise ratio per solution and carry out 5 ADMM iterations with a regularisation parameter of $\rho=5$ using \textsc{sagecal-co} \citep{yatawatta2016}. We also remove the $|\vect{u}|<10$ baselines during this step. We do not use a beam model during calibration, and the flux scale of the visibilities post-calibration is on an apparent scale (i.e. uncorrected for the average beam). Currently, the primary beam model for AARTFAAC-HBA tiles is not implemented in \textsc{sagecal-co}, and efforts are ongoing to utilise the present LOFAR-HBA tile beam model for AARTFAAC-HBA calibration. Therefore, to avoid notable differential primary beam effects, all six sources chosen for this intermediate gain calibration step reside within a $\sim 7^{\circ}$ radius around the NCP. The corrected visibilities are close to the correct flux scale since the primary beam does not change substantially within this radius. 
 \item \label{step:clean} The calibrated visibilities are imaged and deconvolved using the multiscale \texttt{CLEAN} feature of \textsc{wsclean} \citep{offringa2014,offringa2017}, using a cleaning threshold of $0.7\sigma$, with a `Briggs --0.1' weighting-scheme and $|\vect{u}| > 50$ baselines in order to avoid bright degree-scale diffuse emission severely affecting the deconvolution process.
    
 \item \label{step:selfcal} The steps \ref{step:DIcal} and \ref{step:clean} were repeated three times in a "self-cal" loop using the improved calibration model (updated with the clean components) obtained after each iteration. For each iteration, we use an auto-masking threshold \citep{offringa2017} of $8\sigma$, $5\sigma$, and $3\sigma$ respectively, where $\sigma$ is the image rms. The final CLEAN model contains~$4924$ components (delta functions and Gaussians) to represent the compact sources in the field.

 \item \label{step:imaging} We finally image the DI-calibrated visibilities obtained after step \ref{step:selfcal} with \textsc{wsclean} using a `Briggs +0.5' weighting scheme in all subsequent analyses.  

\end{enumerate}

Figure~\ref{fig:dirty_compare} shows single sub-band Stokes $I$ dirty images of the NCP field produced using A12-HBA and LOFAR-HBA data at 122.06~MHz (195.3~kHz bandwidth). The A12-HBA system has a $\sim25$ times larger field of view compared to the LOFAR-HBA system. Galactic large-scale diffuse emission is clearly visible around NCP in the A12-HBA image (see also \citealt{bernardi2010}). Although the diffuse emission is also partly present in LOFAR-HBA, it is not visible because LOFAR-HBA has a much lower short-baseline density and a total absence of baselines shorter than 40~metres ($|\vect{u}| < 15$ or $\ell < 100$ at 122 MHz) compared to A12-HBA.  

\section{Modelling the Diffuse Galactic Emission}\label{sec:modeling}

In this section, we present two approaches to model the diffuse structure around the NCP as seen in figure \ref{fig:dirty_compare}. In the first approach, we use multiscale CLEAN deconvolution with \textsc{wsclean} \citep{offringa2017} to model the diffuse structure. In the second method, we use an orthonormal set of Hermite polynomial basis functions called ``shapelets" \citep{refregier2003,yatawatta2011} to model the diffuse structure. 

\begin{figure*}
    \centering
    \includegraphics[width=0.9\textwidth]{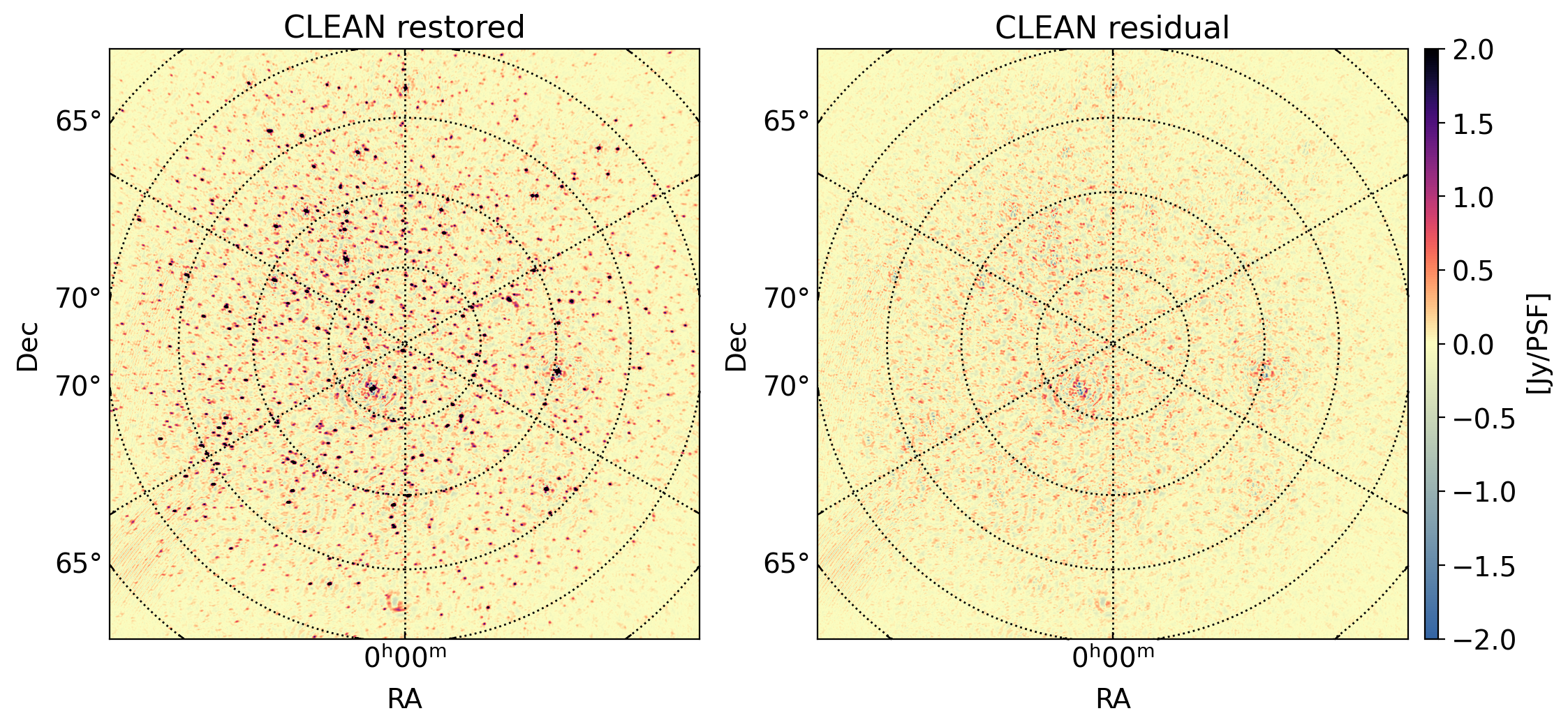}
    \caption{The left panel shows the Stokes $I$ CLEAN restored image of the NCP at 122 MHz after CLEAN deconvolution, produced using a `Briggs --0.1' weighting scheme and with $\vect{u} > 50$ baselines. The right panel shows the Stokes $I$ CLEAN residual image after subtraction of the CLEAN model with 4924 components from the data. The residuals are consistent with the confusion noise.}
    \label{fig:clean_compare}
\end{figure*}

\begin{figure}
    \centering
    \includegraphics[width=\columnwidth]{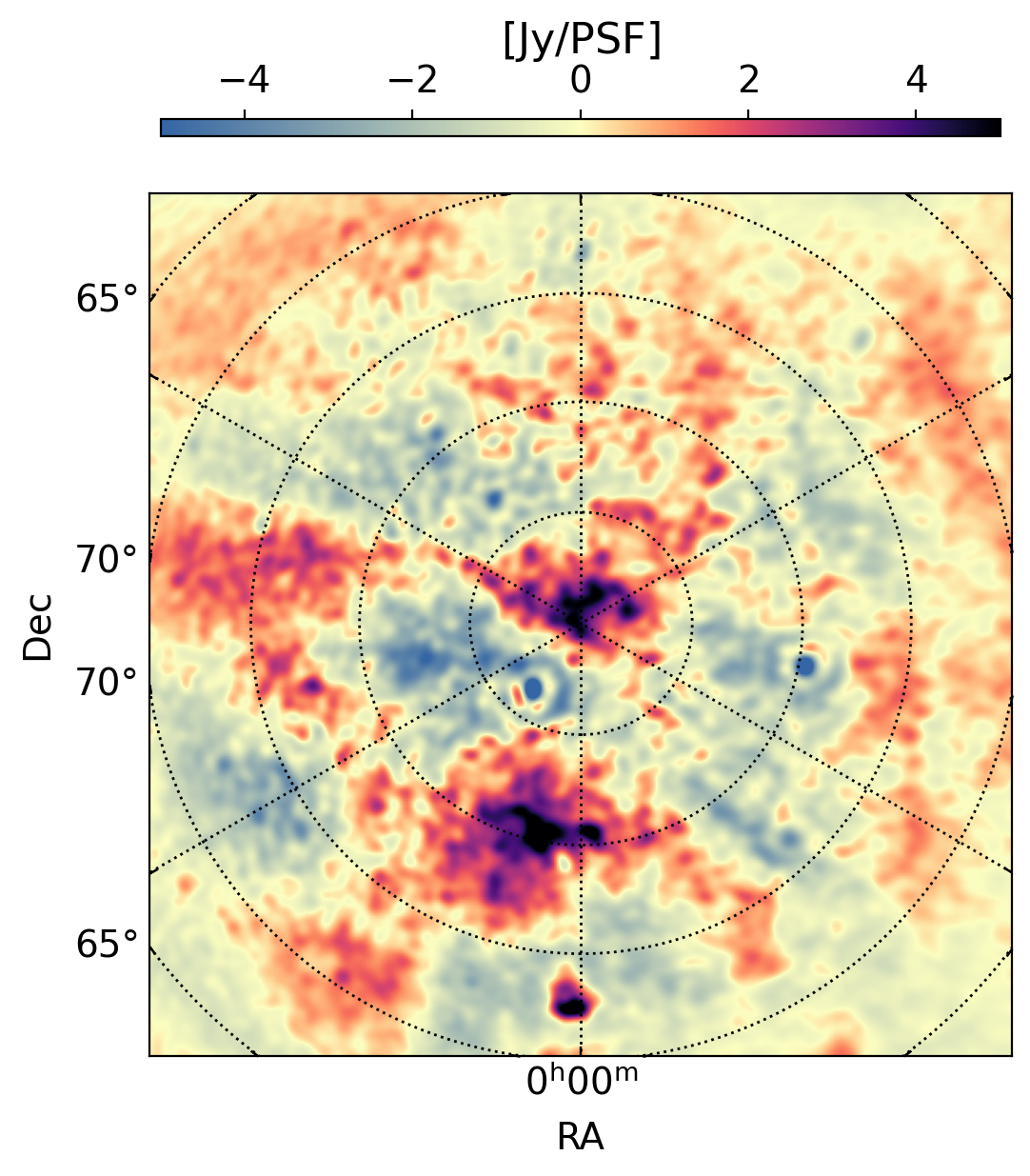}
    \caption{The Stokes $I$ dirty image of the residual visibilities after the subtraction of 4924 CLEAN model components. The image was produced using a 'Briggs +0.5' weighting scheme (which gives more weight to short baselines to better visualise the diffuse emission), $\vect{u} \leq 120$ baseline range, and a Gaussian taper of $30\arcmin$.}
    \label{fig:residual}
\end{figure}

\begin{figure}
    \centering
    \includegraphics[width=\columnwidth]{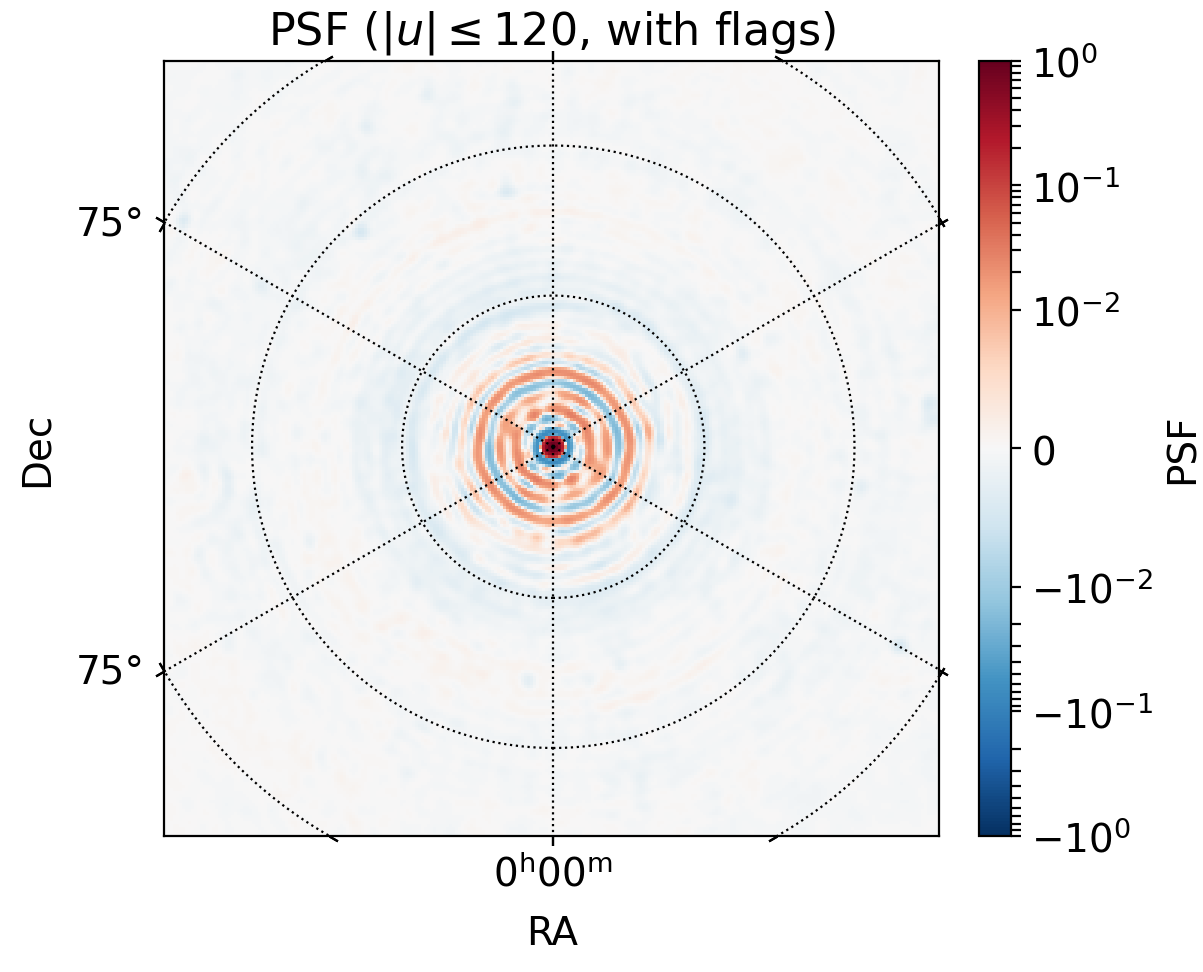}
    \caption{The Point Spread Function of the A12-HBA array towards the NCP produced using 'Briggs +0.5' weighting scheme and $\vect{u} \leq 120$ baselines.}
    \label{fig:A12-HBA-psf}
\end{figure}

\subsection{Removal of compact sources}

Before proceeding with the diffuse foreground modelling, the bright compact sources need to be removed from the map. We use the CLEAN component model obtained during the gain calibration (see step \ref{step:selfcal} in Section \ref{sec:cal_img}) to subtract 4924 components from the data. Figure \ref{fig:clean_compare} shows Stokes $I$ image of the NCP field before and after the subtraction of the CLEAN model. The images are produced using 'Briggs --0.1' weighting scheme with $\vect{u} > 50$ baselines. We observe that subtracting the CLEAN model removes most of the compact sources, leaving only faint residuals of the order of a few hundred mJy rms. The images do not visually reveal signatures of diffuse emission because of the baseline cut used during the imaging process. We also see that the subtraction of some of the brightest sources leaves negative or ring-like artefacts near their source location. Factors such as imperfect calibration and imperfect source modelling are the likely cause. These artefacts can be further mitigated by using DD-calibration to subtract the bright sources. We plan to employ the DD-calibration for the subtraction of bright sources using several directions in future analyses. For the purpose of this work, i.e. analysing the properties of the diffuse foregrounds, these minor artefacts are not important. 

We derive an estimate of the confusion noise for AARTFAAC from the classical confusion limit ($\sigma_c$) using the  relation from \cite{vanhaarlem2013}:
\begin{equation}
\sigma_c = 30 \left( \dfrac{\theta}{1^{\arcsec}} \right)^{1.54} \left( \dfrac{\nu}{74\,\textrm{MHz}} \right)^{-0.7} \ [\mu\textrm{Jy}\,\textrm{beam}^{-1}], 
\end{equation}
where $\theta$ is the angular resolution (FWHM), and $\nu$ is the observation frequency. A12-HBA has an angular resolution of $\sim 7\arcmin$ at 122\,MHz, yielding a confusion limit of $\sigma_c \sim 230$\,mJy\,beam$^{-1}$. The standard deviation of the residuals, after cleaning, for the inner $10^{\circ}$ region of the field is $\sim220$\,mJy. These two values are consistent with each other, assuming that the primary beam correction is small for this inner $10^{\circ}$ region. In the future, we plan a deeper multi-frequency CLEAN by combining sub-bands spanning a wide frequency range along with an improved direction-dependent calibration step that includes an HBA-tile beam model. Figure \ref{fig:residual} shows the Stokes $I$ image of the residual visibilities after subtraction of the 4924 component CLEAN model, produced using $\vect{u} \leq 120$ baselines. The corresponding A12-HBA PSF (shown in figure~\ref{fig:A12-HBA-psf}) is well behaved with few-percent side-lobe levels within $\sim3^{\circ}$ radius and close to zero elsewhere. We now model this residual diffuse emission using the two methods mentioned above.

\begin{figure*}
    \centering
    \includegraphics[width=\textwidth]{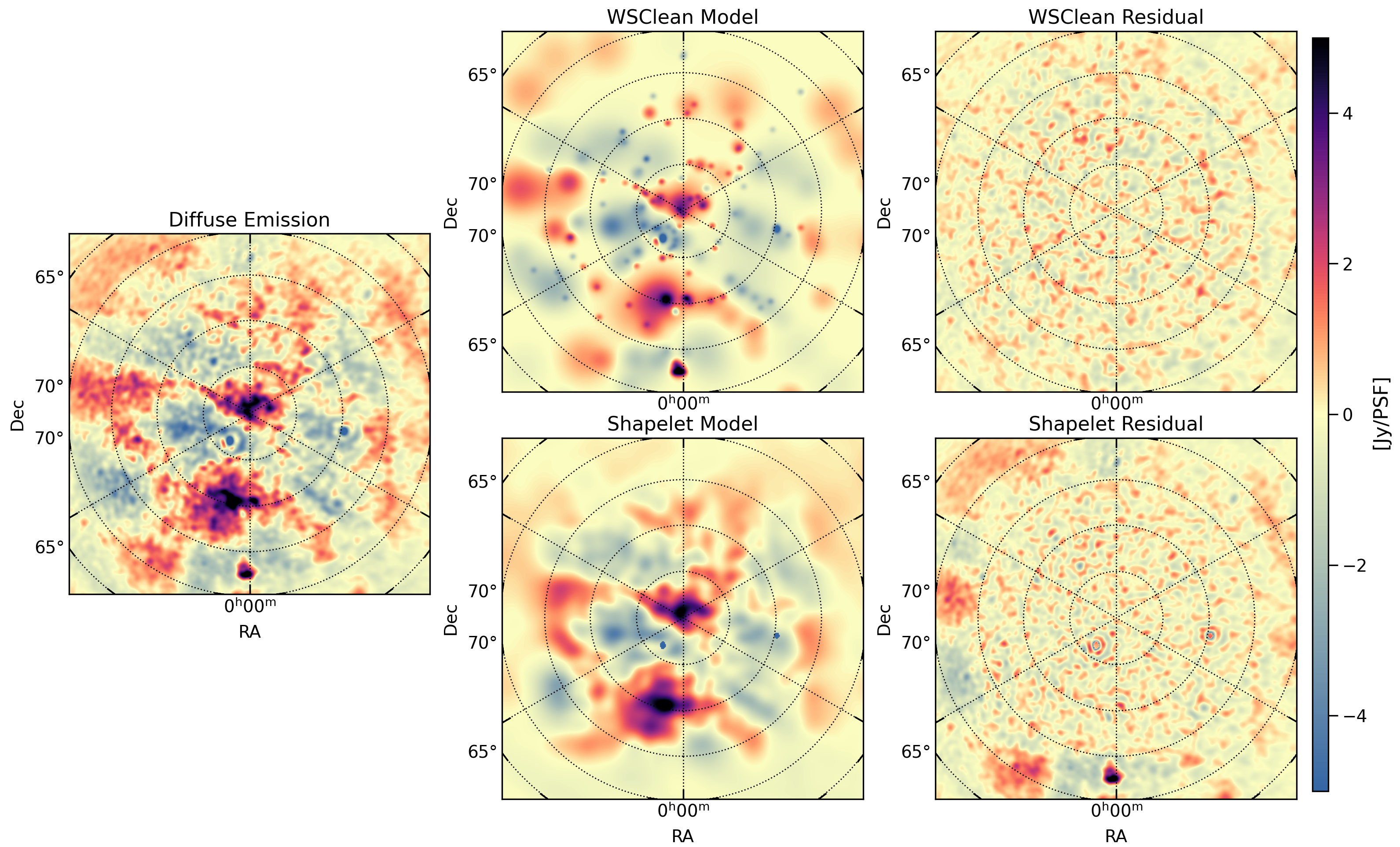}
    \caption{Top row: Stokes $I$ dirty image before subtraction of the diffuse structure (left panel), the diffuse model obtained with multiscale CLEAN (middle panel), and the residual Stokes $I$ image after subtraction of the CLEAN model (right panel). Bottom row: same as the top row but with shapelet decomposition technique to model the diffuse structure. These images were produced using 'Briggs +0.5' and $\vect{u} \leq 120$ baselines with a Gaussian taper of $30\arcmin$.}
    \label{fig:method_compare}
\end{figure*}

\subsection{Modelling with multiscale CLEAN}\label{subsec:multiscale_clean_modeling}
We use the multiscale deconvolution algorithm implemented in \textsc{wsclean} \citep{offringa2017} to model the diffuse structure observed in figure \ref{fig:residual}. We perform the deconvolution with the following settings: `Briggs +0.5' weighting scheme, auto-mask of $3\sigma$, major iteration cleaning gain (mgain) of $0.8$, and a clean stopping threshold of $0.6\sigma$. The `Briggs' weighting scheme with a threshold greater than zero gives more weight to short baselines and better reveals the diffuse emission. Figure \ref{fig:method_compare} (top row) shows the Stokes $I$ image of the field before and after subtracting the diffuse foreground model obtained from multiscale CLEAN. We observe that multiscale CLEAN is able to model the extended emission at different scales using Gaussians and point sources, adequately capturing the diffuse flux. The algorithm also models the compact residuals (both positive and negative) initially missed by the compact model subtraction step. After the subtraction of the diffuse model, the residual appears to be devoid of diffuse structure, with a standard deviation of $\sim 500$~mJy inside the inner region of the image. Note that the CLEAN algorithm is sensitive to image weighting and sampling function applied during the deconvolution process. A possible improvement in the model may be achieved by tuning the image weighting, scales used for deconvolution, and adding more data in the $uv$-plane to cover the (unmodeled) spatial scales with lower sensitivity. For the specific case presented in the paper, we think that multiscale CLEAN, in its current implementation, is a good choice for modelling large-scale diffuse emission. 

\subsection{Modelling with Shapelets}\label{subsec:shapelet_modeling}

\cite{yatawatta2011} showed that the use of orthonormal basis functions e.g.\ Cartesian shapelets \citep{refregier2003} or Prolate Spheroidal Wave Functions (PWSF) \citep{slepian1961,landau1961} in radio interferometric image deconvolution, provides improved image fidelity and larger dynamic range. In the image domain, $(l,m)$, shapelet basis functions can be written as \citep{yatawatta2011}:
\begin{eqnarray}
\phi_{n_1,n_2}(l,m,\beta)  =   \dfrac{1}{2^{n_1 + n_2}\,\piup\beta^2 n_1!\, n_2!}\, H_{n_1}(l/\beta)\,H_{n_2}(m/\beta) \nonumber \\  \times \exp{[-(l^2 + m^2)/2\beta^2]},
\end{eqnarray}

where the functions $H_{n_1}$ and $H_{n_2}$ are the Hermite polynomials of order $n_1$ and $n_2$  being integer values, the value of $\beta$ is the model scale factor. We demonstrate the use of Cartesian shapelets to model the diffuse structure observed in A12-HBA data using the \textsc{shapelet\_gui}\footnote{developed by Sarod Yatawatta.  \url{https://github.com/SarodYatawatta/shapeletGUI}} tool. We use $25\times25$ basis functions for the shapelet decomposition using L1-norm regularisation. We optimise the model scale during the process. Figure \ref{fig:method_compare} (bottom row) presents the Stokes $I$ images before and after shapelet model subtraction, as well as the shapelet model of the diffuse structure obtained from the decomposition. We note that the shapelet model captures the large-scale diffuse structure within the primary beam. The residuals within the primary beam have a standard deviation of $\sim 540$~mJy,  which is approximately the same as residuals after CLEAN diffuse model subtraction. These residuals are most likely the unmodeled compact sources and sources below the confusion noise. However, the diffuse structure outside the primary beam is rather poorly modelled. This is mainly due to the finite support of the basis functions and using a single shapelet model with a large number of basis functions leading to a sub-optimised scale factor ($\beta$,  width of the basis functions). This results in an inadequate performance of the shapelet model in capturing the structure outside the primary beam. The performance of the shapelet decomposition modelling is expected to be enhanced by optimising the number of basis functions used for the decomposition and their scale factor ($\beta$). Additionally, using multiple shapelet models to capture the diffuse structure in different regions in the image space and using primary-beam corrected images is expected to improve the shapelet model. 

\subsection{Comparing the two diffuse emission modelling approaches}

From the visual inspection of the two models in figure \ref{fig:method_compare}, we find that, in this particular setup, the multiscale CLEAN deconvolution method captures compact sources and diffuse structures at a range of angular scales. On the other hand, the shapelet decomposition captures the structure on large scales significantly better. However, it is less effective (without hugely increasing the number of basis functions and shapelets) in modelling compact sources (apparent from the lack of compact structure in the shapelet model in figure~\ref{fig:method_compare}). We infer that in the current implementation, both multiscale CLEAN and shapelet decomposition are able to model the large-scale (order of several degrees) diffuse Galactic emission. However, the shapelet model performs poorly outside the primary beam and is inefficient at capturing the structure on small scales ($<1$~degree). A possible improvement in the CLEAN model may be achieved by tweaking the image weighting scheme and the scales used in deconvolution. Similarly, optimising the number of basis functions, their scale factor, and using multiple shapelets are expected to enhance the shapelet model. However, the CLEAN deconvolution provides the added benefit of fitting a spectrum to capture the spectral behaviour of the diffuse structure, which is currently not the case with shapelet decomposition. The inclusion of spectral indices (per basis function) in the shapelet model is currently under development. For future analyses, we plan to improve the analysis and modelling process by adding more data over a wide range of frequencies, and tweaking various parameters described above. We also plan to explore a hybrid approach to build the sky model by combining the two methods, where the CLEAN model is used to capture small-scale structures and compact sources, and shapelets are used to represent the large-scale diffuse structures. The more complete diffuse emission model will consist of information on a wide range of spatial scales and spectral information. This model can also be utilised to improve the calibration and foreground removal (using DD-calibration) of the LOFAR-EoR and AARTFAAC-HBA observations.  

\begin{figure}
    \centering
   \includegraphics[width=\columnwidth]{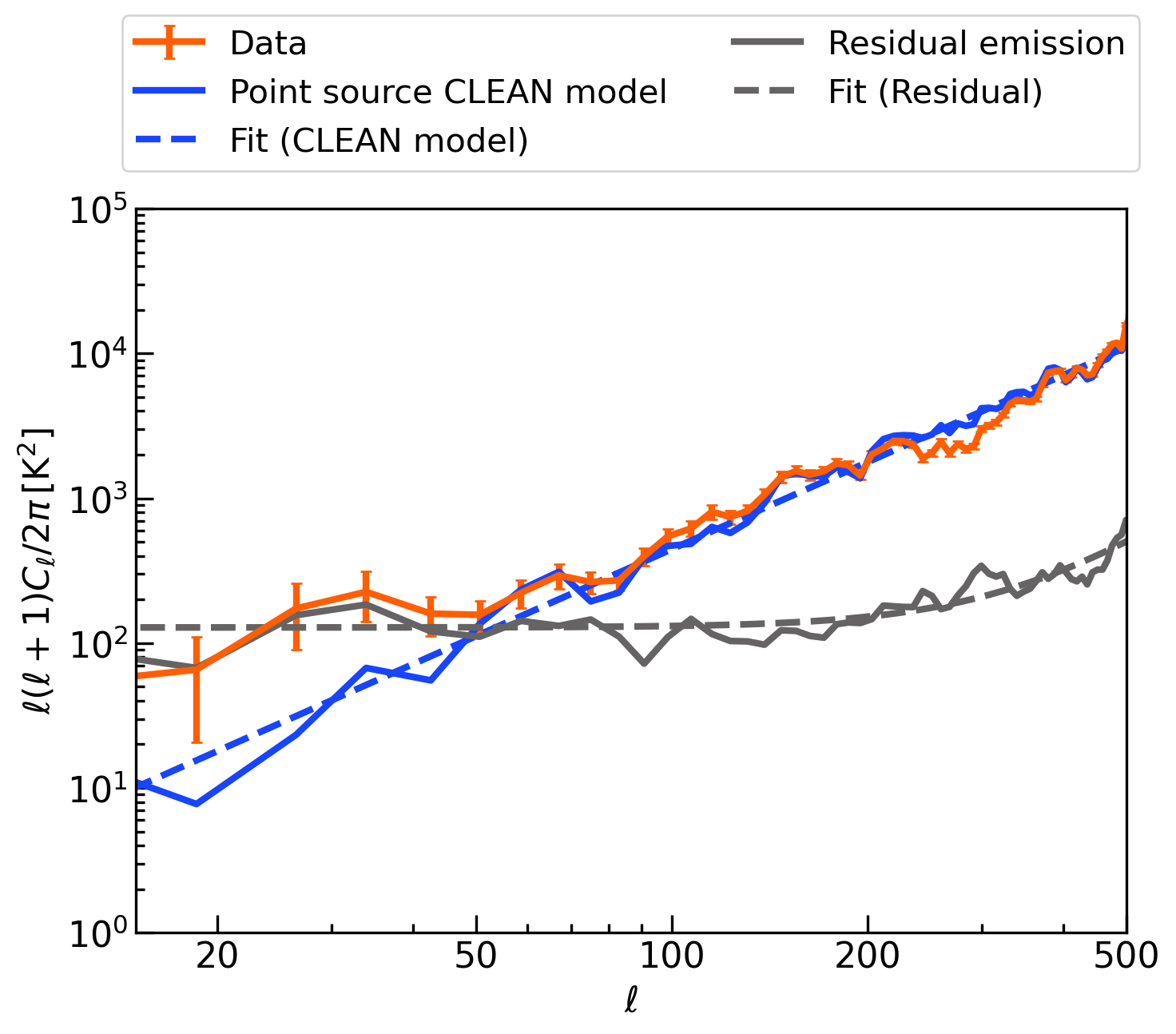}
    \caption{Comparison of the angular power spectrum ($\Delta_{\ell}^2$) of the A12-HBA Stokes $I$ image before and after subtraction of the 4924 component CLEAN model. The solid orange, blue, and grey curves correspond to the emission in the field (data before point source CLEAN model subtraction), the point source CLEAN model, and the residual emission, respectively. The dashed blue and grey curves represent the best-fit profiles of the point source CLEAN model ($\Delta_{\textrm{CLEAN}}^2(\ell)$) and the residual emission ($\Delta_{\textrm{residual}}^2(\ell)$), respectively. The error bars on data correspond to the $2\sigma$ uncertainty due to  sample variance.}
    \label{fig:CL_compact_sub}
\end{figure}

\section{The Angular Power Spectrum}\label{sec:angular_PS}

The angular power spectrum is a commonly used metric to study the spatial properties of the foreground emission. We use the angular power spectrum, $C_\ell$, to describe the observed diffuse emission in the NCP field. The angular power spectrum $C_\ell$ is defined as \citep[see e.g.,][]{seljak1997,bernardi2009}:
\begin{equation}
C_{\ell} = \dfrac{1}{N_\ell} \sum_{\vect{l}} \tilde{I}(\vect{l}) \tilde{I}^*(\vect{l}), 
\end{equation}
where $\tilde{I}(\vect{l})$ is the spatial Fourier transform of an image $I$ (in units of Kelvins), $\vect{l} = 2\piup \vect{u}$ is a two-dimensional vector and Fourier dual to the angular coordinate vector in image space, $N_{\ell}$ is the number of samples in a given azimuthal bin (annulus) of width $\delta|\vect{l}| = 8.0$, the scalar multipole moment $\ell$ is the average of $|\vect{l}|$ in a given azimuthal bin. The multipole moment $\ell$ is related to the angular scale $\Theta$ (Half-Width Half Maximum) in degrees as $\ell = 180^{\circ}/\Theta$. We use the Stokes~$I$ images to estimate $C_\ell$. The images are Fourier transformed along the spatial axes using Fast Fourier Transform (FFT). Because the shapelet modelling performs poorly outside the primary beam,  we multiply the images with a spatial Tukey (tapered cosine) window before the FFT to suppress the emission outside the $15^\circ$ radius. We normalise the Fourier transform of the images with the PSF to remove the effect of the weighting scheme applied during imaging. Note that we do not apply a primary-beam correction during the estimation of $C_\ell$, nor correct for curvature of the sky. We expect the A12-HBA primary beam (or the LOFAR-HBA tile beam) to be fairly regular on the sky-region within the primary beam FWHM. Any beam taper is absorbed in the model and should not affect the single sub-band analysis. The sky-curvature manifests itself as the $w$-term in interferometric imaging. The A12 array is highly co-planar, with the maximum $w$-term only being $\sim1-2\%$ ($\sim6$ in units of wavelength) of the longest baseline. Additionally, $w$-stacking algorithm of \textsc{wsclean} in the current analysis uses 48 $w$-layers to correct for the wide-field/sky curvature related effects. We, therefore, expect that both these effects are limited and of second order in the current analysis.

Figure \ref{fig:CL_compact_sub} presents the angular power spectrum $\Delta^2_\ell \equiv \ell(\ell+1)C_{\ell}/2\piup$ of the Stokes~$I$ image before and after subtraction of the 4924 components of the point source CLEAN model. We do not subtract the noise bias, however, it is expected to be a small fraction of the observed power. We observe that the power on smaller scales (large $\ell$ values) is dominated by compact sources and is mostly removed after subtraction of the CLEAN model. However, power on $\ell < 60$ remains intact after model subtraction. Moreover, the angular power spectrum of the residual emission is flat over a wide range of multipole moments, ranging from $\ell = 20$ to $200$, which corresponds to angular scales of $\Theta = 0.9-10^{\circ}$. To further quantify the point source CLEAN model and residual emission power spectra, we fit the corresponding angular power spectra respectively with the following functions:
\begin{equation}\label{eqn:best_fit_model}
{\rm Point~source~CLEAN~model:~} \Delta^2_{\ell,{\rm CLEAN}} = A \left(\dfrac{\ell}{\ell_0} \right)^{\alpha},     
\end{equation}
\begin{equation}\label{eqn:best_fit_residual}
{\rm Residual~emission:~} \Delta^2_{\ell,{\rm Residual}} =  A^\prime + B^\prime \left(\dfrac{\ell}{\ell_0} \right)^{\alpha^\prime},
\end{equation}
where $A$, $\alpha$, $A^\prime$, $B^\prime$, and $\alpha^\prime$ are the free parameters for the CLEAN model fit, and the residual emission fit, respectively. We choose $\ell_0 = 180$ which correspond to an angular scale $\Theta = 1^{\circ}$. The best-fit parameter values for the two models are listed in Table~\ref{tab:fit_params}. The data and the point source CLEAN model follow power-law like behaviour on $\ell \gtrsim 60$. Fitting the angular power spectrum of the point source CLEAN Model with equation~\ref{eqn:best_fit_model} yields a best fit power-law index $\alpha = 1.99 \pm 0.09$, which is consistent with the angular power spectrum of unresolved and unclustered sources i.e., $\Delta_{\ell}^2 \propto \ell^2$ \citep{Tegmark1996}. This suggests that the compact sources begin dominating the power on baselines $\vect{u} \gtrsim 10$ or angular scales of $\Theta \lesssim 3^{\circ}$ ($\ell \gtrsim 60$). The angular power spectrum ($\Delta_{\ell}^2$) of the residual emission is flat within $20\lesssim \ell \lesssim 200$, suggesting that $C_{\ell} \propto \ell^{-2}$. This is consistent with the angular power spectrum of the Galactic diffuse emission, $C_{\ell}\propto \ell^{-2.2}$ observed by \cite{bernardi2009}. However, the power on $\ell > 200$ is possibly dominated by thermal noise and low level emission below confusion noise, and starts to follow a power-law like behaviour. As revealed from the fit, the residual diffuse emission has a brightness temperature variance of $\Delta^2_{\ell=180} = (145.64 \pm 13.61)\, {\rm K}^2$ at 122\,MHz on angular scales of 1~degree, and is consistent with $C_{\ell}\propto \ell^{-2.0}$ in $20\lesssim \ell \lesssim 200$ range. Using observations of the NCP with Westerbork Synthesis Radio Telescope (WSRT), \cite{bernardi2010} reported $\Delta^2_{\ell=120} \sim 14\,{\rm K}^2$ at 150~MHz, which corresponds to $\Delta^2_{\ell=120} \sim 40\,{\rm K}^2$ at 122~MHz when extrapolated with the spectral index of~$-2.55$. Additionally, \cite{offringa2022} observed $\Delta^2_{\ell=120} \sim 3\,{\rm K}^2$ for the field centred at ``coldest patch" of the northern Galactic hemisphere \citep{kogut2011}, and $\Delta^2_{\ell=120} \sim 10\,{\rm K}^2$ for another field $\sim 10^\circ$ away from the ``coldest patch" at 140 MHz with LOFAR. Whereas, we observe $\Delta^2_{\ell=120} \sim 130\,{\rm K}^2$ in our analysis, that is around four~times the power observed by \cite{bernardi2010}. We hypothesise that the high power observed in our analysis is possibly due to the limited bandwidth of AARTFAAC data ($195$~kHz) used in this study compared to the analyses in \cite{bernardi2010} ($40$~MHz) and \cite{offringa2022} ($48$~MHz). Moreover, we do not apply any noise correction (instrumental and confusion) contrary to \cite{bernardi2010}. We defer the latter for the future analyses.

\begin{table}
    \centering
    \caption{Best-fit Parameters}
    \label{tab:fit_params}
    \begin{tabular}{ll} 
        \hline                 
        \textbf{Parameter} & \textbf{Best-fit value} \\
        \hline
        \multicolumn{2}{c}{CLEAN point sources model} \\
        \hline        
        $A$     & $1438.75 \pm 111.08$ \\ 
        $\alpha$ & $1.99 \pm 0.09$ \\
        \hline
        \multicolumn{2}{c}{Residual emission model} \\
        \hline
        $A^\prime$ & $128.16 \pm 13.45$ \\
        $B^\prime$ & $17.48  \pm 2.08$ \\    
        $\alpha^\prime$ & $3.00 \pm 0.47$ \\
        \hline
    \end{tabular}
\end{table}

\begin{figure}
    \centering
   \includegraphics[width=\columnwidth]{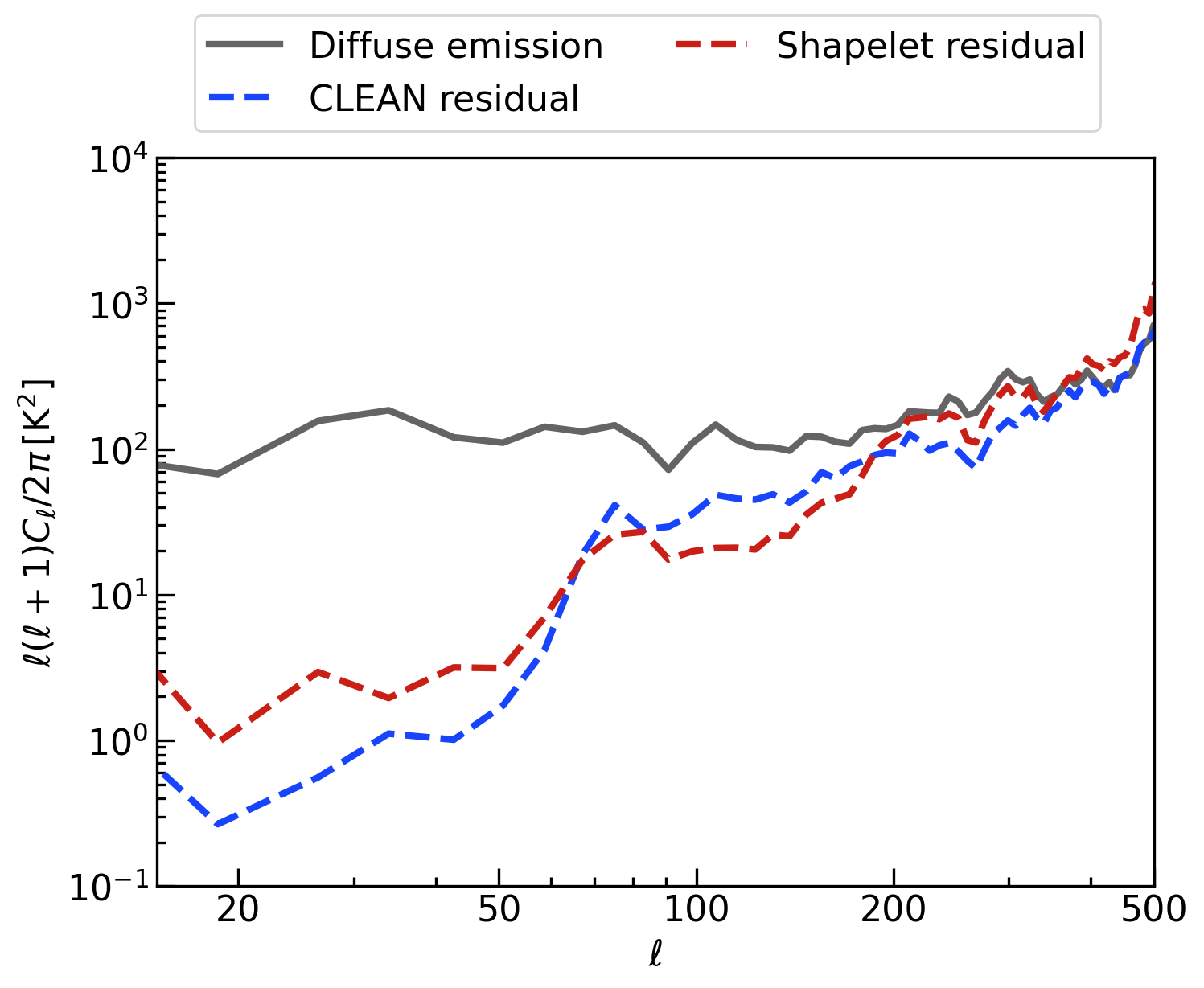}
    \caption{The angular power spectra ($\Delta_{\ell}^2$) of the diffuse emission (grey curve, same as the grey curve in figure~\ref{fig:CL_compact_sub})s, CLEAN residual after subtraction of the CLEAN diffuse model (blue curve), and shapelet residual after subtraction of the shapelet diffuse model (orange curve) shown in figure \ref{fig:method_compare}.}
    \label{fig:CL_compare}
\end{figure}

\subsection{Power spectra from diffuse foreground modelling methods}

The angular power spectrum is also a useful metric to compare the two foreground modelling methods, i.e., multiscale CLEAN and shapelet decomposition, that we demonstrated in Section \ref{sec:modeling}. We use the images shown in figure \ref{fig:method_compare} for the two methods to determine the corresponding angular power spectra $\Delta_{\ell}^2$. We use the same methodology as mentioned in the previous section to determine $C_{\ell}$. Figure \ref{fig:CL_compare} shows the angular power spectra $\Delta_{\ell}^2$ of the residuals after subtraction of CLEAN and shapelet diffuse models obtained from the two methods. We observe that the multiscale CLEAN method leads to lower residuals than the shapelet model on $\ell < 70$ modes ($\Theta \lesssim 2.3^{\circ}$). However, the shapelet model produces slightly lower residuals on $70\lesssim \ell < 200$ compared to the multiscale CLEAN. The residuals on $\ell \gtrsim 200$ behave similarly to the data, probably due to the noise in the data. Note that the two methods (and the residuals obtained from these) compared in this analysis use a very specific setup. A detailed investigation to study and quantify finer differences between the two methods will be performed in future analyses.

\section{Summary and future work}\label{sec:summary_futurework}

In this work, we presented the first-ever wide-field images obtained with the LOFAR AARTFAAC-HBA system in \texttt{A12} mode. In particular, we find strong degree-scale diffuse Galactic Stokes $I$ radio emission at 122\,MHz within a $\sim$20-degree radius field around the North Celestial Pole, which is one of the primary windows of the LOFAR EoR project \citep{yatawatta2013,mertens2020}. We have compared two different methods for modelling this diffuse emission namely, multiscale CLEAN deconvolution and shapelet decomposition. We use angular power spectrum to quantify the behaviour of different foreground components, viz., point sources and diffuse emission. The main results of this work are summarised as follows:

\begin{itemize}
    \item  Stokes $I$ radio emission, as seen by the LOFAR AARTFAAC-HBA system around the NCP at 122\,MHz on baselines shorter than 120 wavelengths, is dominated by large-scale diffuse emission. The angular power spectrum of the emission in the field is dominated by the point sources on scales smaller than 3~degrees ($\ell\gtrsim60$). After subtraction of the point source CLEAN model, the residuals are dominated by the diffuse emission on scales larger than a degree ($\ell\lesssim200$). This diffuse emission, with$~10^4$ times more power than the 21-cm signal, can have a considerable impact on the calibration of any radio-interferometric instrument (e.g., LOFAR, MWA, HERA, NenuFAR, LWA, and SKA) if this emission is not part of the sky model during instrumental gain calibration (or filtered out before calibration) or if their baselines are not excluded during calibration.
    \item We show that, with our particular setup, multiscale CLEAN can model the small and intermediate scales well, but shows a slightly worse performance on intermediate scales than the shapelet decomposition, which captures the large-scale diffuse emission well, but it is incapable of modelling the emission on scales smaller than several degrees.
    \item The diffuse emission has a brightness temperature variance of $\Delta^2_{\ell=180} = (145.64 \pm 13.61)\, {\rm K}^2$ at 122\,MHz on angular scale of 1~degree, and is consistent with $C_{\ell}\propto \ell^{-2.0}$ in $20 \lesssim \ell \lesssim 200$ range.
    
\end{itemize}

The analysis in this pilot work is admittedly based only on data from a single sub-band at 122\,MHz. Despite this restriction, for the first time using AARTFAAC-HBA as a wide-field imaging instrument, we have convincingly shown that large-scale diffuse emission dominates the emission in Stokes~$I$ on degree scales and larger. This was previously much harder to assess due to the limited short-baseline coverage of LOFAR in beam-formed mode. In the current analysis, we use the traditional CLEAN deconvolution to model and subtract the compact sources before modelling the diffuse structure. We suspect that artefacts leftover from the deconvolution and subtraction of compact sources may lead to low-level systematics in the diffuse emission models. In future analyses, we plan to implement direction-dependent calibration to subtract the compact sources instead of just using the CLEAN deconvolution. It is expected to mitigate the systematics arising from poorly subtracted compact sources while taking the primary beam and ionospheric effects into account. In addition to this, we plan to include the following improvements in future analyses for accurate modelling of the diffuse emission:
\begin{itemize}
    \item Including an HBA-tile beam model in the gain calibration, which allows us to (i) set an absolute flux scale, (ii) improve the calibration quality, (iii) perform deeper deconvolution during imaging and (iv) improve the modelling of compact sources and diffuse emission.
    
    \item Expand this analysis to a broader frequency range to obtain a complete spatial and spectral model of the diffuse emission.
    
    \item Explore a hybrid approach to model the emission on a wide range of spatial scales by combining multiscale CLEAN to model small-scale compact emission and shapelet decomposition to model large-scale diffuse emission.
    
    \item Extend the analysis to model diffuse polarised emission in Stokes $Q$ and $U$. Wide-field polarization studies will enable us to understand the behaviour and morphology of polarised foregrounds on very large scales and how it instrumentally leaks to Stokes $I$ \citep{asad2015,asad2016,asad2018,nunhokee2017,byrne2021}.
\end{itemize}

We expect that the diffuse foreground model obtained after including the above-described improvements in the analysis should be sufficient in removing the diffuse emission around the NCP on angular scales and frequencies used to obtain the model regardless of the instrument design. The diffuse emission model will be included in the sky-model used in the LOFAR-EoR and NenuFAR-CD calibration pipelines. The improved diffuse model will also be made available for public use in the form of a git repository. The CLEAN and shapelet models obtained in the current analysis may be obtained by submitting a request to the corresponding author.


\begin{acknowledgements}
BKG and LVEK acknowledge the financial support from the European Research Council (ERC) under the European Union’s Horizon 2020 research and innovation programme (Grant agreement No. 884760, "CoDEX"). FGM and MK acknowledge support from a SKA-NL roadmap grant from the Dutch Ministry of OCW. This work was supported in part by ERC grant 247295 ``AARTFAAC" to RAMJW. LOFAR, the Low Frequency Array designed and constructed by ASTRON, has facilities in several countries, owned by various parties (each with their own funding sources), and collectively operated by the International LOFAR Telescope (ILT) foundation under a joint scientific policy. The authors gratefully acknowledge the feedback from the anonymous referee that helped improve and clarify this work.
\\
This research made use of publicly available software developed for LOFAR and AARTFAAC telescopes. Listed below are software packages used in the analysis: \textsc{aartfaac2ms} (\url{https://github.com/aroffringa/aartfaac2ms}), \textsc{aoflagger} (\url{https://gitlab.com/aroffringa/aoflagger}), \textsc{sagecal-co} (\url{http://sagecal.sourceforge.net/}), and \textsc{wsclean} (\url{https://gitlab.com/aroffringa/wsclean}). The analysis also relies on the Python programming language (\url{https://www.python.org}) and several publicly available python software modules: \textsc{astropy} (\url{https://www.astropy.org/}; \citealt{astropy2013}), \textsc{matplotlib} (\url{https://matplotlib.org/}), \textsc{scipy} (\url{https://www.scipy.org/}) and \textsc{numpy} (\url{https://numpy.org/}).
\end{acknowledgements}

%
%
\bibliographystyle{aa}
\bibliography{diffuse_FG}

\end{document}